\documentclass[prd,nofootinbib,preprint,superscriptaddress]{revtex4}

\usepackage{amsmath, amssymb, amsthm, graphicx, epsfig, fancyhdr,epsfig, slashed}

\usepackage{tikzsymbols}
\usepackage{natbib}
\usepackage{float}

\usepackage{tikz,xcolor,hyperref}

\definecolor{lime}{HTML}{A6CE39}
\DeclareRobustCommand{\orcidicon}{
	\begin{tikzpicture}
	\draw[lime, fill=lime] (0,0) 
	circle [radius=0.2] 
	node[white] {{\fontfamily{qag}\selectfont \tiny ID}};
	\draw[white, fill=white] (-0.0625,0.095) 
	circle [radius=0.007];
	\end{tikzpicture}
	\hspace{-2mm}
}

\foreach \x in {A, ..., Z}{\expandafter\xdef\csname orcid\x\endcsname{\noexpand\href{https://orcid.org/\csname orcidauthor\x\endcsname}
			{\noexpand\orcidicon}}
}

\newcommand{\be}{\begin{equation}}
\newcommand{\ee}{\end{equation}}
\newcommand{\bea}{\begin{eqnarray}}
\newcommand{\eea}{\end{eqnarray}}

\begin{document}

\title{Strongly Coupled String-inspired Infinite Derivative Non-local Yang-Mills: Diluted Mass Gap \footnote{This paper is the original version of the article accepted for publication JHEP (2021) with the title "Diluted Mass Gap in Strongly Coupled Non-local Yang-Mills.}}

\author{Marco Frasca\orcidA{}}
\email{marcofrasca@mclink.it}
\affiliation{Rome, Italy}

\author{Anish Ghoshal\orcidB{}}
\email{anish.ghoshal1@protonmail.ch}
\affiliation{INFN - Sezione Roma “Tor Vergata”, Via della Ricerca Scientifica 1, 00133, Roma, Italy}

\begin{abstract}
\textit{We  investigate  the  non-perturbative regimes in the class of non-Abelian theories that have been proposed as an ultraviolet completion of 4-D Quantum Field Theory (QFT) generalizing the kinetic energy operators to an infinite series of higher-order derivatives inspired by string field theory. We  prove  that,  at  the  non-perturbative  level,  the  physical  spectrum  of  the  theory  is  actually corrected  by the  ”infinite  number  of  derivatives”  present  in  the  action.   
We derive a set of Dyson-Schwinger equations in differential form, for correlation functions till two-points, the solution for which are known in the local theory. We obtain that just like in the local theory, the non-local counterpart displays a mass gap, depending also on the mass scale of non-locality, and show that it is damped in the deep UV asymptotically. We point out some possible implications of our result in particle physics and cosmology and discuss aspects of non-local QCD-like scenarios. }

\end{abstract}

\maketitle

\section{Introduction}


Adopting novel perspectives, our understanding of the fundamental nature of reality may reside in the putative finiteness of superstring loop amplitudes via the Polyakov representation \cite{Polchinski:1998rq,Polchinski:1998rr}. This novel formalism invokes the fact that the fundamental constituent to be quantized is an extended non-local object. 


In this context, an infinite derivative non-local approach was motivated starting from string field theory \cite{Moffat:1990jj, Evens:1990wf, Tomboulis:1997gg, Moffat:2011an, Tomboulis:2015gfa,Kleppe:1991rv,sft1,sft2,sft3,padic1,padic2,padic3,Frampton-padic,Tseytlin:1995uq,marc,Siegel:2003vt,Calcagni:2013eua,Modesto:2011kw,Modesto:2012ga,Modesto:2015foa,Modesto:2017hzl} where attempts where made to address the divergence problem in QFT by generalizing the kinetic energy operators of the Standard Model (SM) to an infinite series of higher order derivatives suppressed by the scale of non-locality ($M$) at which the higher order derivatives come into the picture \cite{Biswas:2014yia}. Such a theoretical construction naturally solves the SM vacuum instability problem  as the $\beta$-functions at the scale of non-locality vanish beyond M, without  introducing any new degrees of freedom in the particle spectrum \cite{Olive:2016xmw,Ghoshal:2017egr}. They have been explicitly shown to be ghost-free \cite{Buoninfante:2018mre}, predicting conformal invariance in the UV, trans-planckian scale transmutation and dark matter phenomenology, 
leading to a new directions of UV-completion of 4D Quantum Field Theories, valid and perturbative up to infinite energy scales \cite{Ghoshal:2018gpq,Buoninfante:2018gce} \footnote{Theories of gravity with infinite derivatives were studied in  Ref.\cite{Biswas:2011ar,Giacchini:2018wlf,Burzilla:2020utr}, particularly for singularities, such as black hole singularities~\cite{Biswas:2011ar,Biswas:2013cha,Frolov:2015bia,Frolov:2015usa,Koshelev:2018hpt,Koshelev:2017bxd,Buoninfante:2018xiw,Cornell:2017irh,Buoninfante:2018rlq,Buoninfante:2018stt,Abel:2019zou,Buoninfante:2020ctr} and cosmological singularities~\cite{Biswas:2005qr,Biswas:2006bs,Biswas:2010zk,Biswas:2012bp,Koshelev:2012qn,Koshelev:2018rau} and in context to inflationary cosmology \& predictions in the CMB,
\cite{Koshelev:2016vhi,SravanKumar:2018dlo,Koshelev:2020fok,Koshelev:2020xby,Koshelev:2020foq}. For SUSY see Refs. \cite{Gama:2017ets,Gama:2020pte}}. 

Strongly coupled sectors in such a scenario was studied in a scalar field theory case and was found to predict diluted mass gap in the UV, with M determining the corrections to the mass spectrum arising from the infinite sets of higher-derivatives present in the theory. Alongside with detailed investigations of N-point correlation functions were performed in Ref.\cite{Frasca:2020jbe}. The purpose of this paper is to extend our analysis to Yang-Mills case and understand the Green's functions in the otherwise non-perturbative regions of the interactions. We will see that the mass gap developed in the local theory gets asymptotically diluted and scale invariance is asymptotically reached in the deep UV.

This paper is organized as follows: In Sec.\ref{Sec1}, we introduce the non-local Yang-Mills theory and fix the formalism. In Sec.\ref{Sec2}, we obtain the Dyson-Schwinger equations for n-point correlation functions (dubbed nP-functions in the following) and obtain the structure of the 2P-function and the mass gap. In Sec.\ref{Sec3}, we put this work in perspective to future possible applications framing better its understanding. 
Finally, in Sec.\ref{Sec6}, we yield the conclusions.

\section{Infinite Derivative SU(N)}
\label{Sec1}

For local non-Abelian SU(N) gauge theory, the Lagrangian includes the gauge boson kinetic term, 
\be
\mathcal{L}_{g}=-\frac{1}{4}  F^{a\mu\nu} F_{a\mu\nu}.
\ee
The trace is over the SU(N) group indices and the field-strength tensor is given by
\be
F_{\mu\nu}^a=\partial_{[\mu}A_{\nu]}^a -gf^{abc}A^b_{\mu}A^c_{\nu}     \ ,
\ee
where the $f^{abc}$ represents the group structure constant and $g$ the coupling that is dimensionless. 
For implementation of the non-local modification, we follow the approach in Refs.~\cite{Ghoshal:2017egr,Ghoshal:2020lfd}. 

It should be pointed out that a common definition in literature is the following \cite{Tomboulis:1997gg,Ghoshal:2020lfd}
\begin{equation}
\label{eq:nlF2}
    L_f = -\frac{1}{4}F^a_{\mu\nu}e^{-f(D^2)}F^{a\mu\nu}.
\end{equation}
This definition is somewhat misleading for the following reason. Let us take
\begin{equation}
    f(D^2)=\frac{D^2}{M^2}
\end{equation}
where $D_\mu^{ab}=\partial_\mu \delta^{ab}-igA^c_{\mu}(T^{c})^{ab}$
is the covariant derivative in the adjoint representation and, as usual, $M$ is a very large mass that sets the scale of non-locality. It should be very large as non-local effects have not been uncovered so far. So, generally speaking, the variation in momenta scale of the $D^2$ will be at most comparable but lesser to $M^2$. Now,
\begin{equation}
    D^2=(\partial_\mu-igT^aA^a_\mu)^2=
    \partial^2-ig\partial^\mu\left(T^aA^a_\mu\right)-igT^aA^a_\mu
    \partial^\mu-g^2T^aT^bA^a_\mu A^{b\mu}.
\end{equation}


By a direct application of the Backer-Campbell-Hausdorff formula (BCH), this yields
\begin{eqnarray}
    e^{-\frac{D^2}{M^2}}&=&e^{-\frac{\Box}{M^2}}e^{-\frac{1}{M^2}\left(-ig\partial^\mu\left(T^aA^a_\mu\right)-igT^aA^a_\mu\partial^\mu-g^2T^aT^bA^a_\mu A^{b\mu}\right)}\times \\
    &&e^{-\frac{1}{2M^4}\left[\Box,-ig\partial^\mu\left(T^aA^a_\mu\right)-igT^aA^a_\mu\partial^\mu-g^2T^aT^bA^a_\mu A^{b\mu}\right]}\times\ldots, \nonumber
\end{eqnarray}
that implies that the commutators are all higher orders with respect to first two exponentials entering as powers of the field. Indeed, there are two kind of approximations that we introduce at this stage. Firstly, we are assuming that the gauge field $A^a_\mu$ is very small with respect to the non-local mass scale. Secondly, we assume that also its space-time variations happen on scale smaller than that of the non-local mass scale. This will imply that a gauge change will not affect the action in a sensible way keeping invariance in a proper limit. 

In order to explicitly evaluate this in the strong coupling limit, let us consider
\be
 e^{-\frac{D^2}{M^2}}=\sum_{n=0}^\infty\frac{(-1)^n}{n!}\left(\frac{D^2}{M^2}\right)^n
 =
 \sum_{n=0}^\infty\frac{(-1)^n}{n!M^{2n}}\left(\partial^2-ig\partial^\mu\left(T^aA^a_\mu\right)-igT^aA^a_\mu
    \partial^\mu-g^2T^aT^bA^a_\mu A^{b\mu}\right)^n.
\ee
In the strong coupling limit, the theory acquires a scale mass $\mu$ and the background solution we start from for SU(N) has the form $A_\mu^a\propto \mu(2/Ng^2)^\frac{1}{4}f(x,\mu,Ng^2)$, where $f(x,\mu,Ng^2)$ is a bounded function (see \cite{Frasca:2015yva} for the local case). This implies that we get a power series in $\mu/M$ and terms like $M^{-1}\partial_\mu$. We are granted that gauge invariance is preserved wherever $\mu/M\ll 1$ and gradient terms can be retained. At the same time, the strong coupling limit $Ng^2\gg 1$ holds consistently and helps to enforce our approximation. Therefore, our techniques can be safely applied. Such an ordering is essential for what follows where we assume that fields are large locally but not with respect to the non-local mass scale $M$. So, adding further fields to preserve gauge invariance, at this stage, would imply to keep higher order terms in our approximation as the scale $\mu/M\ll 1$ will determine their importance.

This argument is consistent with Ref. \cite{Tomboulis:1997gg} where all the non-local factors in the gauge field propagator are properly evaluated with $\Box$ rather than $D^2$ but in that case, the author works in the weak coupling limit. Our approximation is just dual to it, for the local limit, and consistent as well.


The exponential term is introduced by the non-local modification and the Lagrangian includes
an infinite series of higher dimensional operators that are all suppressed by the non-local scale $M$. 
As a result, their contribution can be largely ignored at energies lower than $M$. 
In other words, the conventional Lagrangian is reproduced in the limit of $M \to \infty$. 
We take the metric convention with ${\rm diag}(+1,-1,-1,-1)$ to implement our procedure for UV completion upon the Wick rotation.


\subsection{Gauge Field Re-definition}

Now, we show that we can treat eq.(\ref{eq:nlF2}) 
as for the scalar field. Let us define
\be
  {\hat A}_\mu^a=e^{-\frac{1}{2}f(\Box)}A_\mu^a.
\ee
The idea with this change of variables is to eliminate the exponential non-local factor from the kinetic term just as happens for scalar field theory. Indeed, our work will parallel the one already presented in \cite{Frasca:2020jbe}.


We will arrive at the Lagrangian
\begin{eqnarray}
\label{eq:nlYM1}
\mathcal{L}_{f}&=&\frac{1}{2}{\hat A}_\mu^a(\Box\eta^{\mu\nu}-\partial^\mu\partial^\nu){\hat A}_\nu^a \\
&&-\frac{g}{4}f^{abc}e^{-f(\Box)}\left[e^{\frac{1}{2}f(\Box)}\left(\partial_\mu {\hat A}^a_\nu-\partial_\nu {\hat A}_\mu^a\right)\left(e^{\frac{1}{2}f(\Box)}{\hat A}^{b\mu}
e^{\frac{1}{2}f(\Box)}{\hat A}^{c\nu}\right)\right] \nonumber \\
&&-\frac{g}{4}f^{abc}e^{-f(\Box)}\left[e^{\frac{1}{2}f(\Box)}{\hat A}^{b\mu}
e^{\frac{1}{2}f(\Box)}{\hat A}^{c\nu}e^{\frac{1}{2}f(\Box)}\left(\partial_\mu {\hat A}^a_\nu-\partial_\nu {\hat A}_\mu^a\right)\right] \nonumber \\
&&-\frac{g^2}{4}f^{abc}f^{cde}e^{-f(\Box)}\left[e^{\frac{1}{2}f(\Box)}{\hat A}^{b\mu}e^{\frac{1}{2}f(\Box)}{\hat A}^{c\nu}
e^{\frac{1}{2}f(\Box)}{\hat A}^{d}_\mu e^{\frac{1}{2}f(\Box)}{\hat A}^{e}_\nu\right] \nonumber \\
&&+j_\mu^ae^{\frac{1}{2}f(\Box)}{\hat A}^{a\mu},
\end{eqnarray}
where we added an arbitrary source term $j_\mu^a$ that will be useful in the following. This is similar in the way the non-local scalar field theory is formulated. The main difference is the multiplication of the interaction part by the non-local factor $e^{-f(\Box)}$.




%



The non-Abelian ghost and gauge-fixing Lagrangians are given by
\be
\mathcal{L}_{ghost}=-\bar{c}^ae^{-f(\Box)}(\partial^\mu D_{\mu}^{ab})c^b,
\label{NLghost}
\ee
and
\begin{equation}
\mathcal{L}_{g-f} = 
\frac{1}{2 \xi}{\hat A}_{\mu}^ae^{-f(\Box)}\partial^{\mu}\partial^{\nu}{\hat  A}_{\nu}^{a}, 
\label{NLgauge}
\end{equation}
where $\xi$ is the gauge fixing parameter. In order to have consistency with the standard gauge fixing procedure, we choose the entire function $e^{-f(\Box)}$.
The standard result is obtained in the local limit of $M \to \infty$.

Similarly as we have done for the gauge field, also the ghost field can be redefined as
\begin{equation}
    c^a=e^{-\frac{1}{2}f(\Box)}{\hat c}^a.
\end{equation}
This will yield
\begin{equation}
    \mathcal{L}_{ghost} =
    -{\bar{\hat c}^a}
    \partial^\mu\left(\partial_\mu \delta^{ab}-ige^{\frac{1}{2}f(\Box)}{\hat A}^c_{\mu}(T^{c})^{ab}\right)
    {\hat c}^b+{\bar\eta}^ae^{\frac{1}{2}f(\Box)}{\hat c}^a+e^{\frac{1}{2}f(\Box)}{\bar{\hat c}^a}\eta.
\end{equation}
Also in this case, we added arbitrary source terms $\eta^a$ and ${\bar\eta}^a$.



Finally, our Lagrangian will be given by
\be
\label{eq:fullL}
\mathcal{L}=\mathcal{L}_{f}+\mathcal{L}_{g-f}+\mathcal{L}_{ghost}.
\ee



\section{N-point Correlation Functions}
\label{Sec2}

The aim of this section is to obtain the Dyson-Schwinger equations for the correlation functions of the non-local Yang-Mills theory. Then, we solve them using a generalized form of the mapping theorem \cite{Frasca:2007uz,Frasca:2009yp}. This theorem permits to match the classical solutions of a local scalar field theory on the classical solutions of a local Yang-Mills theory. We will show how this works also for the non-local case. This will provide the spectrum of the non-Abelian theory for the latter case.






\subsection{Classical infinite derivative Yang-Mills theory}

The aim of this section is to show that there is a way to solve the classical Yang-Mills theory that successfully parallels the Dyson-Schwinger approach. This is a current expansion, assuming the gauge potentials as functions of the currents themselves. This was initially presented in \cite{Frasca:2013tma} for the scalar field but it can be applied straightforwardly to any field theory.

We are working fixing the gauge to Landau-Lorenz. The main reason to do this is that, when we will quantize the theory, this choice will make our computations simpler. By no means, this imply that the same computations cannot be done in another or more convenient gauge. Our main result will be the spectrum of the theory that is a gauge-invariant observable and can be obtained in lattice computations at least for the local case.

We can write down the equations of motion for the Yang-Mills field in the form
\begin{eqnarray}
\label{eq:CYM}
\Box A_\mu^a+gf^{abc}
e^{-\frac{1}{2}f(\Box)}\partial_\nu\left[e^{\frac{1}{2}f(\Box)}{\bar A}^{b}_\mu
e^{\frac{1}{2}f(\Box)}A^{c\nu}\right]+&& \nonumber \\
gf^{abc}e^{-\frac{1}{2}f(\Box)}\left[ e^{\frac{1}{2}f(\Box)}A^{b\nu}
e^{\frac{1}{2}f(\Box)}(\partial_\mu A^{c}_\nu-\partial_\nu A^{c}_\mu)\right]+&& \nonumber \\
g^2f^{abc}f^{cde}e^{-\frac{1}{2}f(\Box)}
\left[e^{\frac{1}{2}f(\Box)}A^{b\nu}e^{\frac{1}{2}f(\Box)}A^{d}_\nu
e^{\frac{1}{2}f(\Box)}A^{e}_\mu\right]&=&e^{\frac{1}{2}f(\Box)}j^a_\mu,
\end{eqnarray}
where we have no ghost field, this will enter when we will quantize the theory. We have removed the hat from the fields being the context clear. Then, assuming we are in a strong coupling limit, we can evaluate a strong coupling series by a functional Taylor series in the currents as
\bea
A_\mu^a[j]&=&A_\mu^a[0]+
\int d^4x_1\left.\frac{\delta A_\mu^a}{\delta j_\nu^b(x_1)}\right|_{j=0}j_\nu^b(x_1) \nonumber \\
&&+\frac{1}{2!}
\int d^4x_1d^4x_2\left.\frac{\delta^2A_\mu^a}{\delta j_\nu^b(x_1)j_\lambda^c(x_2)}\right|_{j=0}j_\nu^b(x_1)j_\lambda^c(x_2)+O(j^3).
\eea
The coefficients, that play a similar role to the correlation functions of the quantum theory that we will see in a moment, can obtained from eq.(\ref{eq:CYM}) by deriving it successively with respect to the currents and solving the corresponding PDEs one obtains. We do not give here this set of equations because are not of interest for what follows.

Anyway, we point out that, In order to solve this hierarchy of equations, there is a class of solutions for the local Yang-Mills theory that maps on a $\phi^4$ theory. In principle, this permits to solve all the hierarchy of the Dyson-Schwinger equations \cite{Frasca:2015yva}. Besides, it gives the spectrum of the theory in very close agreement with lattice computations \cite{Frasca:2017slg}. Then, it is legit to guess that we can have such solutions at the leading order that are properly corrected by non-local effects, if we refer to the mass scale $M$ being very high. This mapping was firstly proposed in Refs. \cite{Frasca:2007uz,Frasca:2009yp}.


The technique is the following \cite{Frasca:2019ysi}. Let us introduce a set of symbols: $\eta-$symbols for SU(N) (this can be possibly extend to the SO(N) case in a straightforward way. It should be seen for other groups.). For the sake of simplicity, we work out the case for SU(2) where the proof of their existence is straightforward. Indeed, in this case, they can be defined as 
\cite{Smilga:2001ck,Frasca:2019ysi}
\begin{equation}
\eta_\mu^a=((0,1,0,0),(0,0,1,0),(0,0,0,1)),
\end{equation}
that yields
\begin{equation}
\eta_\mu^1=(0,1,0,0),\ \eta_\mu^2=(0,0,1,0),\ \eta_\mu^3=(0,0,0,1),
\end{equation}
that implies $\eta_\mu^a\eta^{a\mu}=3$. This easily generalizes to SU(N) as \begin{equation}
\label{eq:eta1}
\eta_\mu^a\eta^{a\mu}=N^2-1.
\end{equation}
Similarly, by generalizing the SU(2) case,
\begin{equation}
\eta_\mu^a\eta^{b\mu}=\delta_{ab},
\end{equation}
and
\begin{equation}
\eta_\mu^a\eta_\nu^a=\frac{1}{2}\left(g_{\mu\nu}-\delta_{\mu\nu}\right),
\end{equation}
being $g_{\mu\nu}$ the Minkowski metric and $\delta_{\mu\nu}$ the identity tensor. Then, we choose the solutions as for the local case in the form
\begin{equation}
    A_\mu^a(x)=\eta_\mu^a\phi(x)
\end{equation}
and put this into the action. This will work exactly in the Lorenz gauge. 
%
%
that we fix to carry on the quantum theory otherwise we will get asymptotic corrections \cite{Frasca:2007uz,Frasca:2009yp} in the limit of a strong coupling.

\subsection{Partition function}

Give the Lagrangian in Eqn.(\ref{eq:fullL}), the partition function is given by
\begin{equation}
Z[j,{\bar\eta},\eta]=\int D[A]D[c]D[{\bar c}]
e^{-\int d^4x\mathcal{L}_f-\int d^4x\mathcal{L}_{g-f}-\int d^4x\mathcal{L}_{ghost}}.
\end{equation}
We will refer to it when we will do averaging in quantum field theory. We are working in Euclidean metric.



\subsection{Dyson-Schwinger equations}


We use the Bender-Milton-Savage method \cite{Bender:1999ek}, described in Appendix A, to obtain the set of Dyson-Schwinger equations for our case. Then, we will approach these equations choosing a set of solutions mapped from a scalar field. This will yield in the end the main result of the paper that is, the spectrum of the theory.

We start from the Lagrangian given in Eqn.(\ref{eq:nlYM1}). 
After we have chosen the Landau-Lorenz gauge aiming at quantization, as already said, 
we obtain the motion equations for the gauge fields given by
\begin{eqnarray}
\Box A_\mu^a+gf^{abc}
e^{-\frac{1}{2}f(\Box)}\partial_\nu\left[e^{\frac{1}{2}f(\Box)}{\bar A}^{b}_\mu
e^{\frac{1}{2}f(\Box)}A^{c\nu}\right]+&& \nonumber \\
gf^{abc}e^{-\frac{1}{2}f(\Box)}\left[ e^{\frac{1}{2}f(\Box)}A^{b\nu}
e^{\frac{1}{2}f(\Box)}(\partial_\mu A^{c}_\nu-\partial_\nu A^{c}_\mu)\right]+&& \nonumber \\
g^2f^{abc}f^{cde}e^{-\frac{1}{2}f(\Box)}
\left[e^{\frac{1}{2}f(\Box)}A^{b\nu}e^{\frac{1}{2}f(\Box)}A^{d}_\nu
e^{\frac{1}{2}f(\Box)}A^{e}_\mu\right]+&& \nonumber \\
gf^{abc}e^{\frac{1}{2}f(\Box)}\bar{c}^b\partial_\mu c^c&=&e^{\frac{1}{2}f(\Box)}j^a_\mu,
\end{eqnarray}
and for the ghost field is
\be
-\Box c^a +gf^{abc}\left(e^{\frac{1}{2}f(\Box)}A_\mu^b\right)\partial^\mu c^c=e^{\frac{1}{2}f(\Box)}\eta^a.
\ee

For our computations to follow, we point out that the propagator in the Landau gauge can be written as \cite{Peskin:1995ev}
\begin{equation}
\label{eq:GLG}
    G_{\mu\nu}^{ab}(x,y)=\delta_{ab}\left(\eta_{\mu\nu}-\frac{\partial_\mu\partial_\nu}{\Box}\right)G_2(x,y).
\end{equation}

Our aim will be to get the structure of the 2P-function for the non-local Yang-Mills theory and the mass gap, and study its nature. These are presented in Appendix B.


\subsection{Solution of the Dyson-Schwinger equations}

We can try to get a solution to the Dyson-Schwinger set of equations for the infinite-derivative Yang-Mill theory in a similar way this is done for the local theory \cite{Frasca:2015yva}. The idea is to select a set of solutions and work out all the results (background field method). Anyway, it is possible to start with a different set of solutions, e.g. a Yang-Wu monopole \cite{Wu:1967vp}, but in this way the results obtained for the local theory are quite different and closed form solutions could not possibly be found. The Dyson-Schwinger set of equations, till 2-P functions, is derived in Appendix B. In order to solve them, starting with the 1P-functions, we look for a solution mapped on a scalar field by taking \cite{Frasca:2015yva} 
\be
\label{eq:phi_map}
G_{1\mu}^a(x)=\eta_\mu^a\phi(x).
\ee
Using the properties of the $\eta$-symbols and the fact that one can take the 1P-functions, $P_1^a(x)=0$ and $K_2^{ab}(0)=0$, we get for eq.(\ref{eq:G1j0}) 
\be
\label{eq:G1phi}
\Box\phi(x)+
Ng^2e^{-\frac{1}{2}f(\Box)}
\left[
2G_2(0)e^{f(\Box)}\phi(x)+
\left(e^{\frac{1}{2}f(\Box)}\phi(x)\right)^3
\right]=0.
\ee
This is very similar to the result we obtained for the scalar field in \cite{Frasca:2020jbe}. The most important change arises from the mass shift due to quantum effects that, in this case, is given by
\be
\delta m^2=2Ng^2G_2(0).
\ee
The factor 2 arises just by the algebraic properties of the gauge group and can change depending on it.

Next, we can consider the 2P-function by applying the mapping with the scalar field in eq.(\ref{eq:phi_map}) and the 2P-function in the Landau gauge as in eq.(\ref{eq:GLG}). For reasons of completeness, we need to report the mapped equation. One gets
\bea
\label{eq:G2phi}
\delta_{ah}\left(\eta_{\mu\lambda}-\frac{\partial_\mu\partial_\lambda}{\Box}\right)\Box G_2(x,y)
\nonumber \\
+gf^{abc}e^{-\frac{1}{2}f(\Box)}\partial^\nu\left[
e^{\frac{1}{2}f(\Box)}G_{3\mu\nu\lambda}^{bch}(0,y)+
e^{\frac{1}{2}f(\Box)}\delta_{bh}\left(\eta_{\mu\lambda}-\frac{\partial_\mu\partial_\lambda}{\Box}\right)G_2(x,y)\times
\right.
\nonumber \\
\left.
e^{\frac{1}{2}f(\Box)}\eta_\nu^c\phi(x)
+e^{\frac{1}{2}f(\Box)}\eta_\mu^b\phi(x)
e^{\frac{1}{2}f(\Box)}
\delta_{ch}\left(\eta_{\nu\lambda}-\frac{\partial_\nu\partial_\lambda}{\Box}\right)G_2(x,y)
\right]- \nonumber \\
gf^{abc}e^{-\frac{1}{2}f(\Box)}
\left[ 
e^{\frac{1}{2}f(\Box)}\partial^\nu G_{3\mu\nu\lambda}^{bch}(0,y)+
e^{\frac{1}{2}f(\Box)}
\delta_{bh}\left(\eta_{\mu\lambda}-\frac{\partial_\mu\partial_\lambda}{\Box}\right)\partial^\nu e^{\frac{1}{2}f(\Box)}G_2(x,y)
\eta_\nu^c e^{\frac{1}{2}f(\Box)}\phi(x)
\right.\nonumber \\
\left.
e^{\frac{1}{2}f(\Box)}\eta_\mu^b\partial^\nu\phi(x)
\delta_{ch}\left(\eta_{\nu\lambda}-\frac{\partial_\nu\partial_\lambda}{\Box}\right)e^{\frac{1}{2}f(\Box)}G_2(x,y)
\right]-\nonumber \\
gf^{abc}e^{-\frac{1}{2}f(\Box)}
\left[
e^{\frac{1}{2}f(\Box)}\partial_\mu G_{3\nu\lambda}^{bch\nu}(0,y)+
e^{\frac{1}{2}f(\Box)}\delta_{bh}\left(\eta_{\nu\lambda}-\frac{\partial_\nu\partial_\lambda}{\Box}\right)\partial_\mu G_2(x,y)
\eta^{c\nu}e^{\frac{1}{2}f(\Box)}\phi(x)+
\right.
\nonumber \\
\left.
e^{\frac{1}{2}f(\Box)}\delta_{ch}\left(\eta_{\lambda}^\nu-\frac{\partial^\nu\partial_\lambda}{\Box}\right)\partial_\mu G_2(x,y)
\eta_\nu^{b}e^{\frac{1}{2}f(\Box)}\partial_\mu\phi(x)
\right]+\nonumber \\
g^2f^{abc}f^{cde}e^{-\frac{1}{2}f(\Box)}
\left[
e^{\frac{1}{2}f(\Box)}G_{3\mu\nu\lambda}^{(j)bdh}(0,y)e^{\frac{1}{2}f(\Box)}\eta^{\nu e}e^{\frac{1}{2}f(\Box)}\phi(x)
\right.
\nonumber \\
e^{\frac{1}{2}f(\Box)}\delta_{bd}\left(\eta_{\mu\nu}-\frac{\partial_\mu\partial_\nu}{\Box}\right)G_2(0)
e^{\frac{1}{2}f(\Box)}\delta_{eh}\left(\eta_{\lambda}^\nu-\frac{\partial_\lambda\partial^\nu}{\Box}\right)G_2(x,y)
+e^{\frac{1}{2}f(\Box)}\partial^\nu G_{4\mu\nu\lambda}^{bdeh\nu}(0,0,y)+
\nonumber \\
e^{\frac{1}{2}f(\Box)}\delta_{bh}\left(\eta_{\mu\lambda}-\frac{\partial_\mu\partial_\lambda}{\Box}\right)G_2(x,y)
\eta_\nu^d\eta^{e\nu}\left(e^{\frac{1}{2}f(\Box)}\phi(x)\right)^2+
\nonumber \\
e^{\frac{1}{2}f(\Box)}\delta_{dh}\left(\eta_{\nu\lambda}-\frac{\partial_\nu\partial_\lambda}{\Box}\right)G_2(x,y)
\eta_\mu^b\eta^{e\nu}\left(e^{\frac{1}{2}f(\Box)}\phi(x)\right)^2+
\nonumber \\
e^{\frac{1}{2}f(\Box)}\delta_{eh}\left(\eta_{\lambda}^\nu-\frac{\partial^\nu\partial_\lambda}{\Box}\right)G_2(x,y)
\eta_\mu^b\eta^{d}_\nu\left(e^{\frac{1}{2}f(\Box)}\phi(x)\right)^2+
\nonumber \\
e^{\frac{1}{2}f(\Box)}G_{3\mu\lambda}^{beh\nu}(0,y)e^{\frac{1}{2}f(\Box)}\eta_{\nu}^{d}\phi(x)+ 
\nonumber \\
\delta_{be}\left(\eta_{\mu}^\nu-\frac{\partial_\mu\partial^\nu}{\Box}\right)e^{\frac{1}{2}f(\Box)}G_2(0)
\delta_{dh}\left(\eta_{\nu\lambda}-\frac{\partial_\nu\partial_\lambda}{\Box}\right)e^{\frac{1}{2}f(\Box)}G_2(x,y)
\nonumber \\
\left.
e^{\frac{1}{2}f(\Box)}G_{3\nu\lambda}^{deh\nu}(0,y)e^{\frac{1}{2}f(\Box)}\eta_{\mu}^{b}\phi(x)
+\delta_{de}\left(\eta_{\nu}^\nu-\frac{\partial_\nu\partial^\nu}{\Box}\right)e^{\frac{1}{2}f(\Box)}G_2(0)
\delta_{bh}\left(\eta_{\mu\lambda}-\frac{\partial_\mu\partial_\lambda}{\Box}\right)e^{\frac{1}{2}f(\Box)}G_2(x,y)
\right]- \nonumber \\
gf^{abc}e^{\frac{1}{2}f(\Box)}
\left\{
{\bar J}_{2\lambda}^{bh}(x,y)e^{\frac{1}{2}f(\Box)}\left[\partial_\mu P_1^{c}(x)\right]\right.+ \nonumber \\
\left.{\bar P}_1^{b}(x)e^{\frac{1}{2}f(\Box)}\left[\partial_\mu J_{2\lambda}^{ch}(x,y)\right]
+\partial_\mu\left[W_{3\lambda}^{bch}(0,y)\right]
\right\}= \nonumber \\
e^{\frac{1}{2}f(\Box)}\delta^{ah}\eta_{\mu\lambda}\delta^4(x-y).
\eea
This equation can be simplified by noticing that the the terms linear $f^{abc}$ goes to zero, due to anti-symmetry of $f^{abc}$, granting the correct symmetry arising from the Kr\"onecker symbol in the first term. The reason for this is quite simply. One has terms like $f^{abc}\eta_b\delta_{ch}$ but, for the equation to have a non-trivial solution, one must have only the terms with $a=h$ due to the fact that we have a factor $\delta_{ah}$ on other terms. This sets those contributions to zero. We also assume that the contributions coming from $G_3$ and $G_4$ are zero. Finally, we use the solution $P_1^a(x)=0$ introduced in the computations of the 1P-function. This will give in the end
\be
\Box G_2(x,y)+\delta m^2 e^{\frac{1}{2}f(\Box)}G_2(x,y)+3Ng^2\left(e^{\frac{1}{2}f(\Box)}\phi(x)\right)^2e^{\frac{1}{2}f(\Box)}G_2(x,y)=e^{\frac{1}{2}f(\Box)}\delta^4(x-y).
\ee
Similarly, for the ghost 2P-functions one gets
\be
-\Box K_2^{ch}(x,y)=e^{\frac{1}{2}f(\Box)}\delta^{ch}\delta^4(x-y),
\ee
and
\be
-\Box J_2^{ch\nu}(x,y)=0.
\ee
We see that, also for non-local Yang-Mills field theory the ghost decouples in the Landau gauge and has the propagator of a free massless particle. This means that we can also assume $J_2^{ch\nu}(x,y)=0$ consistently.

\subsection{Mass gap}


In this section, we aim to compute the spectrum of the theory. This is a gauge-independent observable that is obtained by the 2P-function that, otherwise, are gauge dependent. It seen in the local case that, indeed, the correct spectrum is obtained by comparison with the lattice data \cite{Frasca:2017slg}. Particularly, in the non-local case, the main result is just the mass gap.



As already seen for the scalar field \cite{Frasca:2020jbe}, due to the non-local effects being small corrections to the local case $\phi_c(x)$, we do a negligible error if we put $\phi(x)\approx\phi_c(x)$ into the equation for the 2P-function being \cite{Frasca:2015yva}
\be
\phi_c(x)=\mu\left(\frac{2}{Ng^2}\right)^\frac{1}{4}{\rm sn}\left(p\cdot x+\theta\right)
\ee
being $\mu$ and $\theta$ arbitrary integration constants and provided that
\be
p^2=\mu^2\sqrt{\frac{Ng^2}{2}}.
\ee
We can use the equation \cite{Frasca:2020jbe}
\begin{eqnarray}
    \left(e^{\frac{1}{2}f(\Box)}\phi_c\right)^2&=&
   \nonumber \\
     \left[
    \mu\left(\frac{2}{Ng^2}\right)^\frac{1}{4}\frac{2\pi}{K(i)}\sum_{n=0}^\infty(-1)^n\frac{e^{-\left(n+\frac{1}{2}\right)\pi}}{1+e^{-(2n+1)\pi}}
    e^{\frac{1}{2}f\left(-(2n+1)\frac{\pi}{2K(i)}p^2\right)}\right.\times&& \nonumber \\
    \left.\sin\left((2n+1)\frac{\pi}{2K(i)}(p\cdot x+\theta)\right)\right]^2&&
\end{eqnarray}
that, for our aims, takes the form
\begin{eqnarray}
 \left(e^{\frac{1}{2}f(\Box)}\phi_c\right)^2&=&\mu^2
 \left(\frac{2}{Ng^2}\right)^\frac{1}{2}\frac{4\pi^2}{K^2(i)}\times \nonumber \\
&&\left[\frac{e^{-\pi}}{(1+e^{-\pi})^2}e^{f\left(-\frac{\pi^2}{4K^2(i)}p^2\right)}\sin^2\left(\frac{\pi}{2K(i)}(p\cdot x+\theta)\right)+
\right. \nonumber \\
&&\left.\sum_{n=1}^\infty D_n(x)
\right],
\end{eqnarray}
being $D_n$ all the contributions arising from the square of the series with $n>0$. It is an effect of the non-locality to flatten the spectrum of the theory retaining a mass gap that is diluted in the ultraviolet, as already seen for the scalar field. This happens also in this case as we get the equation for the 2P-function
\begin{eqnarray}
&&-\Box G_2(x,y)+3Ng^2\mu^2
 \left(\frac{2}{Ng^2}\right)^\frac{1}{2}\frac{4\pi^2}{K^2(i)}\times \nonumber \\
&&\left\{\frac{e^{-\pi}}{(1+e^{-\pi})^2}e^{f\left(-\frac{\pi^2}{4K^2(i)}p^2\right)}
\left[1-\cos\left(\frac{\pi}{K(i)}(p\cdot x+\theta)\right)\right]
+\sum_{n=1}^\infty D_n(x)
\right\}e^{\frac{1}{2}f(\Box)}G_2(x,y)\nonumber \\
&&+\delta m^2 e^{\frac{1}{2}f(\Box)}G_2(x,y)=e^{\frac{1}{2}f(\Box)}\delta^4(x-y).
\end{eqnarray}
This equation is linear and we can apply a Fourier transform. Anyway, already at this stage, we can note a mass term given by
\be
\label{eq:mg}
\Delta m^2=
\mu^2
 \left(18Ng^2\right)^\frac{1}{2}\frac{4\pi^2}{K^2(i)}
 \frac{e^{-\pi}}{(1+e^{-\pi})^2}e^{f\left(-\frac{\pi^2}{4K^2(i)}p^2\right)}
 +\delta m^2.
\ee
This must be completed by the gap equation
\be
\label{eq:mgge}
\delta m^2=2Ng^2G_2(0)=2Ng^2\int\frac{d^4p}{(2\pi)^4}G_2(p).
\ee
This result agrees quite well with the mass gap presented in \cite{Frasca:2020jbe} for the scalar field, aside from the factor 2 in the gap equation that is 3 for the scalar field and here arises from algebraic properties of the gauge group.

Working as already done for the scalar field, we can write down the structure of the 2P-function as \cite{Frasca:2020jbe}
\be
\label{eq:G2s}
G_2(k)=\frac{e^{\frac{1}{2}f(-k^2)}}{k^2+\Delta m^2e^{\frac{1}{2}f(-k^2)}}\frac{1}{1-\Pi(k)}
\ee
where the function $\Pi(k)$ and the corresponding sign can be properly obtained by an iteration procedure already described in \cite{Frasca:2020jbe}.




By comparison with \cite{Frasca:2020jbe}, we see that the structure of the two-point function in Eqn.(\ref{eq:G2s}) is very similar to the one obtained for the scalar field. This does not come out as a surprise as we used a mapped solution but it is indeed remarkable that all the techniques devised for the local field theory apply here as well yielding absolutely non-trivial results. 

\medskip 

\section{Non-local QCD-like Theory: A Heuristic Discussion}
\label{Sec3}

\subsection{Non-local QCD-like Model}


Adding Fermions to the theory we discussed in this paper implies that it would be interesting to study its infrared limit. Indeed, we can use the 2P-function we obtained in Eqn.(\ref{eq:G2s}) to devise a non-local Nambu-Jona-Lasinio approximation. We can possibly write
\be
\mathcal{L}_{nlNJL}={\bar\psi}(x)e^{\frac{1}{2}f(\Box)}(i{\slashed\partial}-m)\psi(x)+
g^2\int d^4y{\bar\psi}(x)\psi(x) G_2(x,y){\bar\psi}(y)\psi(y)
\ee
where we used the result given in \cite{Ghoshal:2017egr} for the non-local Dirac equation and omitted internal degrees of freedom. Anyway, this kind of model can imply breaking of chiral symmetry and confinement as recently shown \cite{Frasca:2019ysi}. This kind of model can also describe bounded states and all the properties of QCD at very low-energy can be translated to the non-local case. The advantage to have an entire function multiplying the propagator can assure convergence of the integrals, notably for the chiral condensate. A natural cut-off is also granted by the non-local mass scale $M$.

\subsection{Non-local Confinement Criterion}

In a recent paper \cite{Ghoshal:2020lfd}, it was shown that BRST invariance does not change too much with respect to the local case. This is a rather interesting result in view of the Kugo-Ojima criterion for confinement in a Yang-Mills theory without fermions \cite{Kugo:1977zq,Kugo:1979gm}. It was shown recently that this criterion, obtained by BRST invariance, provides the exact beta function for the local theory that is shown to confine \cite{Chaichian:2018cyv}. Therefore, we are in a very similar situation in the non-local case and we could get a beta function, even if approximately, describing the behaviour of the theory on all the energy range providing a confinement proof.


We extended these ideas to the non-local case and we have shown that, in the infrared limit, there is the coupling running to infinity without a Landau pole. This is a different way to obtain confinement to a trivial fixed point. Also, we see that in the integral that determines the Kugo-Ojima confinement criterion, easily extended to the non-local case, the very existence of a non-local mass scale yields UV contributions that, when integrated, cannot be neglected lowering the energy scale. This is expected as such an integral ranges on all the enrgy scale to infinity \cite{fratp}.


\subsection{Asymptotically Diluted Mass Gap}


Considering Eqn.(\ref{eq:mg}) and (\ref{eq:mgge}) in the UV-limit, the mass gap of the non-local Yang-Mills theory becomes diluted in the UV. So, the mass gap becomes even more negligible as the energy increases. This property of the theory can be helpful to understand several phenomena as dark energy understood as quintessence.



\subsection{Toy Model for Dark Matter}


As we have seen, the non-local Yang-Mills theory appears to have a single stable glueball in the spectrum while higher order excitations are displaced in the very far UV. This could be a possible candidate for dark matter. Possible paths to pursue are comprehensively described in \cite{Boddy:2014yra,Boddy:2014qxa}.

\medskip

\section{Conclusion}
\label{Sec6}

We investigated 
non-local Yang-Mills theory
in 4-D with non-Abelian SU(N) gauge interactions and predicted the mass gap and the mass spectrum of the states in the strong coupled regime. Alongside we compared the results with that of the local theory and discussed implications regarding realistic QCD-like scenarios. We summarize the main findings of our paper below:
\begin{itemize}
    \item We particularly studied the \textit{mass gap} generated in non-local infinite-derivative Yang-Mills field theory and showed, as an example, for the case of non-local Gaussian operator, that the non-local scale M is responsible for no extra poles in the propagator even in the non-perturbative regime (see Eqn.(\ref{eq:G2s})). This feature was already speculated while investigating  the scalar non-local field theory in Ref. \cite{Frasca:2020jbe}. 
    \item As in the case of the scalar field, the generated mass gap gets diluted and asymptotically vanishes in the UV in the limit $M \rightarrow \infty$ or becomes very small near the non-local mass scale..
    \item We have shown that, also for non-local Yang-Mills theory, the mass spectrum
    is yielded by a single mass gap with higher excitations displaced far away in the energy scale with respect to the local theory (see Eqn.(\ref{eq:G2s})). 
\end{itemize}


The approach of evaluation mass gap and N-point functions is general enough that can be extended, in principle, to the case for gravity, as gauge theory in the form of non-local gravity as a direction of re-normalizable, ghost-free quantum gravity. Several possibilities involving particle physics and cosmology open up with our obtained results as already shown in Sectns.\ref{Sec3}. which will pursue the avenues of research incorporating proper treatment of quantum gravity and cosmology in future publications.


\section{Acknowledgements}
\label{Asck}

Authors acknowledge Luca Buoninfante for very useful comments and suggestions on the manuscript.

\newpage

\section*{Appendix A: Dyson-Schwinger Equations \& Bender-Milton-Savage Technique}
\label{AppendixA}


In this appendix, we describe the Bender-Milton-Savage technique \cite{Bender:1999ek} to obtain the hierarchy of equations for the nP-functions of a quantum field theory. The main idea is to retain the PDE form for them without ever introducing the vertex functions.

In order to fix the ideas, we work with a single component scalar field having the following generic partition function
\begin{equation}
    Z[j]=\int[D\phi]e^{iS(\phi)+i\int d^4xj(x)\phi(x)}.
\end{equation}
The first equation for the 1P-function is obtained by
\be
\left\langle\frac{\delta S}{\delta\phi(x)}\right\rangle=j(x)
\ee
where
\be
\left\langle\ldots\right\rangle=\frac{\int[D\phi]\ldots e^{iS(\phi)+i\int d^4xj(x)\phi(x)}}{\int[D\phi]e^{iS(\phi)+i\int d^4xj(x)\phi(x)}}
\ee
We complete the procedure by setting $j=0$. 
Then, we derive the above equation dependent on $j$ to obtain the equation for the 2P-function. We point out that the nP-functions are defined by
\be
\langle\phi(x_1)\phi(x_2)\ldots\phi(x_n)\rangle=\frac{\delta^n\ln(Z[j])}{\delta j(x_1)\delta j(x_2)\ldots\delta j(x_n)}.
\ee
Therefore,
\be
\frac{\delta G_k(\ldots)}{\delta j(x)}=G_{k+1}(\ldots,x).
\ee

So, for a $\phi^4$ theory one will have
\be
S=\int d^4x\left[\frac{1}{2}(\partial\phi)^2-\frac{\lambda}{4}\phi^4\right],
\ee
and we have to evaluate
\be
\label{eq:G_1}
\partial^2\langle\phi\rangle+\lambda\langle\phi^3(x)\rangle = j(x).
\ee
This yields
\be
Z[j]\partial^2G_1^{(j)}(x)+\lambda\langle\phi^3(x)\rangle = j(x).
\ee
Using the definition of the 1P-function it is
\be
Z[j]G_1^{(j)}(x)=\langle\phi(x)\rangle.
\ee
Deriving this with respect to $j(x)$ we obtain
\be
Z[j][G_1^{(j)}(x)]^2+Z[j]G_2^{(j)}(x,x)=\langle\phi^2(x)\rangle.
\ee
Another derivation step gives
\be
Z[j][G_1^{(j)}(x)]^3+3Z[j]G_1^{(j)}(x)G_2(x,x)+Z[j]G_3^{(j)}(x,x,x)=\langle\phi^3(x)\rangle.
\ee
This is substituted into eq.(\ref{eq:G_1}) giving
\be
\label{eq:G1_j}
\partial^2G_1^{(j)}(x)+\lambda[G_1^{(j)}(x)]^3+3\lambda G_2^{(j)}(0)G_1^{(j)}(x)+G_3^{(j)}(0,0)=Z^{-1}[j]j(x)
\ee
We realize that, by the effect of renormalization, a mass term appeared.
This, by setting $j=0$, yields the first Dyson-Schwinger equation into differential form
\be
\partial^2G_1(x)+\lambda[G_1(x)]^3+3\lambda G_2(0)G_1(x)+G_3(0,0)=0.
\ee

We derive again eq.(\ref{eq:G1_j}) with respect to $j(y)$ obtaining
\be
\begin{split}
&\partial^2G_2^{(j)}(x,y)+3\lambda[G_1^{(j)}(x)]^2G_2^{(j)}(x,y)+
\nonumber \\
&3\lambda G_3^{(j)}(x,x,y)G_1^{(j)}(x)
+3\lambda G_2^{(j)}(x,x)G_2^{(j)}(x,y)
+G_4^{(j)}(x,x,x,y)=\nonumber \\
&Z^{-1}[j]\delta^4(x-y)+j(x)\frac{\delta}{\delta j(y)}(Z^{-1}[j]).
    \end{split}
\ee
Setting $j=0$, one gets the equation for the 2P-function as
\be
\partial^2G_2(x,y)+3\lambda[G_1(x)]^2G_2(x,y)+
3\lambda G_3(0,y)G_1(x)
+3\lambda G_2(0)G_2(x,y)
+G_4(0,0,y)=
\delta^4(x-y).
\ee
Such a procedure can be iterated to whatever order giving, in principle, all the set of the Dyson-Schwinger hierarchy's equations in PDE form.


\section*{Appendix B: Derivation of the Dyson-Schwinger equations for 1P- and 2P-functions}
\label{AppendixB}

In this appendix we show all the step needed to obtain the Dyson-Schwinger equations for the 1P- and 2P-functions for the non-local Yang-Mills theory with the Bender-Milton-Savage method. After averaging the equations of motion we get
\begin{eqnarray}
\Box G_{1\mu}^{(j)a}+gf^{abc}
e^{-\frac{1}{2}f(\Box)}\left\langle\partial_\nu\left[e^{\frac{1}{2}f(\Box)}{\bar A}^{b}_\mu
e^{\frac{1}{2}f(\Box)}A^{c\nu}\right]\right\rangle+&& \nonumber \\
gf^{abc}e^{-\frac{1}{2}f(\Box)}\left\langle\left[ e^{\frac{1}{2}f(\Box)}A^{b\nu}
e^{\frac{1}{2}f(\Box)}(\partial_\mu A^{c}_\nu-\partial_\nu A^{c}_\mu)\right]\right\rangle&& \nonumber \\
g^2f^{abc}f^{cde}e^{-\frac{1}{2}f(\Box)}
\left\langle\left[e^{\frac{1}{2}f(\Box)}A^{b\nu}e^{\frac{1}{2}f(\Box)}A^{d}_\nu
e^{\frac{1}{2}f(\Box)}A^{e}_\mu\right]\right\rangle+&& \nonumber \\
+gf^{abc}e^{\frac{1}{2}f(\Box)}\left\langle \bar{c}^b\partial_\mu c^c\right\rangle&=&e^{\frac{1}{2}f(\Box)}j^a_\mu,
\end{eqnarray}
and similarly for the ghost
\be
-\Box P_1^{(\eta)a} +gf^{abc}\left\langle\left(e^{\frac{1}{2}f(\Box)}A_\mu^c\right)\partial^\mu c^b\right\rangle=e^{\frac{1}{2}f(\Box)}\eta^a.
\ee
At this stage, we have introduced the 1P-functions
\bea
\label{eq:defs}
G_{1\mu}^{(j)a}(x)&=&Z^{-1}\langle A_\mu^a(x)\rangle \nonumber \\
P_1^{(\eta)a}(x)=&=&Z^{-1}\langle c^a(x)\rangle.
\eea
A similar equation holds also for ${\bar c}^a$ that gives ${\bar P}_1^{(\eta)a}(x)$. Then, we evaluate the averages appearing in these equations as follows. Let us consider the definitions given above written as
\bea
Z[j,\eta,{\bar\eta}]e^{\frac{1}{2}f(\Box)}G_{1\mu}^{(j)a}(x)&=&\langle e^{\frac{1}{2}f(\Box)}A_\mu^a(x)\rangle \nonumber \\
Z[j,\eta,{\bar\eta}]P_1^{(\eta)a}(x)&=&\langle c^a(x)\rangle.
\eea
We have introduced the apexes $(j)$ and $(\eta)$ to remember the explicit dependence on the currents that will be set to zero to the end of computation. Then, we derive one time with respect to $j(x)$ on the first equation to get
\be
\label{eq:f1}
Ze^{\frac{1}{2}f(\Box)}G_{2\mu\nu}^{(j)ab}(x,x)+
Ze^{\frac{1}{2}f(\Box)}G_{1\mu}^{(j)a}(x)e^{\frac{1}{2}f(\Box)}G_{1\nu}^{(j)b}(x)=
\langle e^{\frac{1}{2}f(\Box)}A_\mu^a(x)e^{\frac{1}{2}f(\Box)}A_\nu^b(x)\rangle.
\ee
Therefore, by applying the space derivative $\partial^\nu$, we also have
\be
Ze^{\frac{1}{2}f(\Box)}\partial^\nu G_{2\mu\nu}^{(j)ab}(x,x)+
Ze^{\frac{1}{2}f(\Box)}\partial^\nu G_{1\mu}^{(j)a}(x)e^{\frac{1}{2}f(\Box)}G_{1\nu}^{(j)b}(x)=
\langle e^{\frac{1}{2}f(\Box)}\partial^\nu A_\mu^a(x)e^{\frac{1}{2}f(\Box)}A_\nu^b(x)\rangle.
\ee
This step is important as such averages enter into the equation for the 1P-function. A further derivation of eq.(\ref{eq:f1}) with respect to $j^{c\nu}$ will yield
\bea
Ze^{\frac{1}{2}f(\Box)}G_{2\mu\nu}^{(j)ab}(x,x)e^{\frac{1}{2}f(\Box)}G_1^{(j)\nu c}(x)
+Ze^{\frac{1}{2}f(\Box)}G_{3\mu\nu}^{(j)abc\nu}(x,x,x)+
\nonumber \\
Ze^{\frac{1}{2}f(\Box)}G_{1\mu}^{(j)a}(x)e^{\frac{1}{2}f(\Box)}G_{1\nu}^{(j)b}(x)e^{\frac{1}{2}f(\Box)}G_1^{(j)\nu c}(x)+Ze^{\frac{1}{2}f(\Box)}G_{2\mu}^{(j)ac\nu}(x)e^{\frac{1}{2}f(\Box)}G_{1\nu}^{(j)b}+\nonumber \\
Ze^{\frac{1}{2}f(\Box)}G_{2\nu}^{(j)bc\nu}(x)e^{\frac{1}{2}f(\Box)}G_{1\mu}^{(j)a}(x)=
\langle e^{\frac{1}{2}f(\Box)}A_\mu^a(x)e^{\frac{1}{2}f(\Box)}A_\nu^b(x)e^{\frac{1}{2}f(\Box)}A^{c\nu}(x)\rangle.
\eea
In order to complete the computation, we need also to evaluate the averages for the ghost field. From eq.(\ref{eq:defs}) we write
\be
\label{eq:P1}
Z[j,\eta,\bar\eta]P_1^{(\eta)a}(x)=\langle c^a(x)\rangle.
\ee
Then, we derive this equation with respect to $\partial_\mu$ and then with respect to $\bar\eta$, the first derivative just enters into the equation of motion of the ghost field while the latter is needed to obtain the correlation function. One gets
\be
Z{\bar P}_1^{(\eta)b}(x)e^{\frac{1}{2}f(\Box)}\partial^\mu P_1^{(\eta)a}(x)+Z\partial^\mu K_2^{(\eta)ab}(x,x)=
\langle {\bar c}^b\partial^\mu c^a(x)\rangle.
\ee
In this equation, we have introduced the new 2P-function 
\be
K_2^{(\eta)ab}(x,y)=\frac{1}{Z}\frac{\delta P_1^{(\eta)a}(x)}{\delta \eta^b(y)}.
\ee
Also, we need the following 2P-function
\be
J_{2\mu}^{(\eta,j)ab}(x,y)=\frac{1}{Z}\frac{\delta P_1^{(\eta)a}(x)}{\delta j^{b\mu}(y)}.
\ee
Therefore, by deriving eq.(\ref{eq:P1}) with respect to $j^{b\mu}(x)$, we get
\be
Ze^{\frac{1}{2}f(\Box)}G_{1\mu}^{(j)b}(x)\partial^\mu P_1^{(\eta)a}(x)+Z\partial^\mu J_{2\mu}^{(\eta,j)ab}(x,x)=
\langle A^b_\mu(x)\partial^\mu c^a(x)\rangle.
\ee
Finally, we can collect all these computations to get
\begin{eqnarray}
\label{eq:G1j}
\Box G_{1\mu}^{(j)a}+gf^{abc}
e^{-\frac{1}{2}f(\Box)}\partial^\nu\left[
e^{\frac{1}{2}f(\Box)}G_{2\mu\nu}^{(j)bc}(x,x)+
e^{\frac{1}{2}f(\Box)}G_{1\mu}^{(j)b}(x)e^{\frac{1}{2}f(\Box)}G_{1\nu}^{(j)c}(x)
\right]-&& \nonumber \\
gf^{abc}e^{-\frac{1}{2}f(\Box)}
\left[ 
e^{\frac{1}{2}f(\Box)}\partial^\nu G_{2\mu\nu}^{(j)bc}(x,x)+
e^{\frac{1}{2}f(\Box)}\partial^\nu G_{1\mu}^{(j)b}(x)e^{\frac{1}{2}f(\Box)}G_{1\nu}^{(j)c}(x)
\right]-
&& \nonumber \\
gf^{abc}e^{-\frac{1}{2}f(\Box)}
\left[ 
e^{\frac{1}{2}f(\Box)}\partial_\mu G_{2\nu}^{(j)bc\nu}(x,x)+
e^{\frac{1}{2}f(\Box)}\partial_\mu G_{1\nu}^{(j)b}(x)e^{\frac{1}{2}f(\Box)}G_{1}^{(j)c\nu}(x)
\right]+
&& \nonumber \\
g^2f^{abc}f^{cde}e^{-\frac{1}{2}f(\Box)}
\left[
e^{\frac{1}{2}f(\Box)}G_{2\mu\nu}^{(j)bd}(x,x)e^{\frac{1}{2}f(\Box)}G_1^{(j)\nu e}(x)
+e^{\frac{1}{2}f(\Box)}\partial^\nu G_{3\mu\nu}^{(j)bde\nu}(x,x,x)+
\right.
\nonumber \\
e^{\frac{1}{2}f(\Box)}G_{1\mu}^{(j)b}(x)e^{\frac{1}{2}f(\Box)}G_{1\nu}^{(j)d}(x)e^{\frac{1}{2}f(\Box)}G_1^{(j)\nu e}(x)+e^{\frac{1}{2}f(\Box)}G_{2\mu}^{(j)be\nu}(x,x)e^{\frac{1}{2}f(\Box)}G_{1\nu}^{(j)d}(x)+\nonumber \\
\left.
e^{\frac{1}{2}f(\Box)}G_{2\nu}^{(j)de\nu}(x,x)e^{\frac{1}{2}f(\Box)}G_{1\mu}^{(j)b}(x)
\right]-&& \nonumber \\
gf^{abc}e^{\frac{1}{2}f(\Box)}
\left\{
{\bar P}_1^{(\eta)b}(x)e^{\frac{1}{2}f(\Box)}\left[\partial_\mu P_1^{(\eta)c}(x)\right]+\partial_\mu\left[K_2^{(\eta)bc}(x,x)\right]
\right\}
&=& \nonumber \\
e^{\frac{1}{2}f(\Box)}j^a_\mu,&&
\end{eqnarray}
It is interesting to point out that, if the exponential due to the non-locality would be set to 1 (i.e. we are taking the local limit $M \rightarrow \infty$), we have just recovered the same equation computed for this case in \cite{Frasca:2015yva}.
The same is true for the ghost field that yield instead
\be
-\Box P_1^{(\eta)c} 
-gf^{abc}e^{\frac{1}{2}f(\Box)}G_{1\mu}^{(j)a}(x)\partial^\mu P_1^{(\eta)b}(x)-gf^{abc}\partial^\mu J_{2\mu}^{(\eta,j)ab}(x,x)
=e^{\frac{1}{2}f(\Box)}\eta^c.
\ee
Following the computational step reported here, we get the first equation for the 1P-function by setting all the currents to zero to obtain
\begin{eqnarray}
\label{eq:G1j0}
\Box G_{1\mu}^{a}+gf^{abc}
e^{-\frac{1}{2}f(\Box)}\partial^\nu\left[
e^{\frac{1}{2}f(\Box)}G_{2\mu\nu}^{bc}(x,x)+
e^{\frac{1}{2}f(\Box)}G_{1\mu}^{b}(x)e^{\frac{1}{2}f(\Box)}G_{1\nu}^{c}(x)
\right]-&& \nonumber \\
gf^{abc}e^{-\frac{1}{2}f(\Box)}
\left[ 
e^{\frac{1}{2}f(\Box)}\partial^\nu G_{2\mu\nu}^{bc}(x,x)+
e^{\frac{1}{2}f(\Box)}\partial^\nu G_{1\mu}^{b}(x)e^{\frac{1}{2}f(\Box)}G_{1\nu}^{c}(x)
\right]-
&& \nonumber \\
gf^{abc}e^{-\frac{1}{2}f(\Box)}
\left[ 
e^{\frac{1}{2}f(\Box)}\partial_\mu G_{2\nu}^{bc\nu}(x,x)+
e^{\frac{1}{2}f(\Box)}\partial_\mu G_{1\nu}^{b}(x)e^{\frac{1}{2}f(\Box)}G_{1}^{c\nu}(x)
\right]+
&& \nonumber \\
g^2f^{abc}f^{cde}e^{-\frac{1}{2}f(\Box)}
\left[
e^{\frac{1}{2}f(\Box)}G_{2\mu\nu}^{bd}(x,x)e^{\frac{1}{2}f(\Box)}G_1^{\nu e}(x)
+e^{\frac{1}{2}f(\Box)}\partial^\nu G_{3\mu\nu}^{bde\nu}(x,x,x)+
\right.
\nonumber \\
e^{\frac{1}{2}f(\Box)}G_{1\mu}^{b}(x)e^{\frac{1}{2}f(\Box)}G_{1\nu}^{d}(x)e^{\frac{1}{2}f(\Box)}G_1^{(j)\nu e}(x)+e^{\frac{1}{2}f(\Box)}G_{2\mu}^{be\nu}(x,x)e^{\frac{1}{2}f(\Box)}G_{1\nu}^{d}(x)+\nonumber \\
\left.
e^{\frac{1}{2}f(\Box)}G_{2\nu}^{de\nu}(x,x)e^{\frac{1}{2}f(\Box)}G_{1\mu}^{b}(x)
\right]-&& \nonumber \\
gf^{abc}e^{\frac{1}{2}f(\Box)}
\left\{
{\bar P}_1^{b}(x)e^{\frac{1}{2}f(\Box)}\left[\partial_\mu P_1^{c}(x)\right]+\partial_\mu\left[K_2^{bc}(x,x)\right]
\right\}=0,&&
\end{eqnarray}
We see that, in the local limit, we have recovered the Dyson-Schwinger equation for the 1P-function of the Yang-Mills theory \cite{Frasca:2015yva}. Similarly, for the ghost field we get
\be
-\Box P_1^{c} 
-gf^{abc}e^{\frac{1}{2}f(\Box)}G_{1\mu}^{a}(x)\partial^\mu P_1^{b}(x)-gf^{abc}\partial^\mu J_{2\mu}^{ab}(x,x)
=0.
\ee
and similarly for ${\bar P}_1^c(x)$.
Equations for 1P-functions are fundamental to apply our background field method. Then, given their solutions, we are able to evaluate the 2P-functions and obtain the observables of the theory.

Moving from eq.(\ref{eq:G1j}), we can get the equations for the 2P-functions in the following way.
We derive it with respect to $j^{\lambda h}(y)$ obtaining
\begin{eqnarray}
\label{eq:G2j}
\Box G_{2\mu\lambda}^{(j)ah}(x,y)+gf^{abc}
e^{-\frac{1}{2}f(\Box)}\partial^\nu\left[
e^{\frac{1}{2}f(\Box)}G_{3\mu\nu\lambda}^{(j)bch}(x,x,y)+
e^{\frac{1}{2}f(\Box)}G_{2\mu\lambda}^{(j)bh}(x,y)\times
\right.&&
\nonumber \\
\left.
e^{\frac{1}{2}f(\Box)}G_{1\nu}^{(j)c}(x)+
+e^{\frac{1}{2}f(\Box)}G_{1\mu}^{(j)b}(x)
e^{\frac{1}{2}f(\Box)}G_{2\nu\lambda}^{(j)ch}(x)
\right]-&& \nonumber \\
gf^{abc}e^{-\frac{1}{2}f(\Box)}
\left[ 
e^{\frac{1}{2}f(\Box)}\partial^\nu G_{2\mu\nu\lambda}^{(j)bch}(x,x,y)+
e^{\frac{1}{2}f(\Box)}\partial^\nu G_{2\mu\lambda}^{(j)bh}(x,y)e^{\frac{1}{2}f(\Box)}G_{1\nu}^{(j)c}(x)+
\right.
&& \nonumber \\
\left.
e^{\frac{1}{2}f(\Box)}\partial^\nu G_{1\mu}^{(j)b}(x)e^{\frac{1}{2}f(\Box)}G_{2\nu\lambda}^{(j)ch}(x,y)
\right]-
&& \nonumber \\
gf^{abc}e^{-\frac{1}{2}f(\Box)}
\left[
e^{\frac{1}{2}f(\Box)}\partial_\mu G_{3\nu\lambda}^{(j)bch\nu}(x,x,y)+
e^{\frac{1}{2}f(\Box)}\partial_\mu G_{2\nu\lambda}^{(j)bh}(x,y)e^{\frac{1}{2}f(\Box)}G_{1}^{(j)c\nu}(x)+
\right.&&
\nonumber \\
\left.
e^{\frac{1}{2}f(\Box)}\partial_\mu G_{1\nu}^{(j)b}(x)e^{\frac{1}{2}f(\Box)}G_{2\lambda}^{(j)ch\nu}(x,y)
\right]+
&& \nonumber \\
g^2f^{abc}f^{cde}e^{-\frac{1}{2}f(\Box)}
\left[
e^{\frac{1}{2}f(\Box)}G_{3\mu\nu\lambda}^{(j)bdh}(x,x,y)e^{\frac{1}{2}f(\Box)}G_1^{(j)\nu e}(x)+
\right.&&
\nonumber \\
e^{\frac{1}{2}f(\Box)}G_{2\mu\nu}^{(j)bd}(x,x)e^{\frac{1}{2}f(\Box)}G_{2\lambda}^{(j)\nu eh}(x,y)
+e^{\frac{1}{2}f(\Box)}\partial^\nu G_{4\mu\nu\lambda}^{(j)bdeh\nu}(x,x,x,y)+
\nonumber \\
e^{\frac{1}{2}f(\Box)}G_{2\mu\lambda}^{(j)bh}(x,y)e^{\frac{1}{2}f(\Box)}G_{1\nu}^{(j)d}(x)e^{\frac{1}{2}f(\Box)}G_1^{(j)\nu
e}(x)+&&
\nonumber \\
e^{\frac{1}{2}f(\Box)}G_{1\mu}^{(j)b}(x)e^{\frac{1}{2}f(\Box)}G_{2\nu\lambda}^{(j)dh}(x,y)e^{\frac{1}{2}f(\Box)}G_1^{(j)\nu
e}(x)+&&
\nonumber \\
e^{\frac{1}{2}f(\Box)}G_{1\mu}^{(j)b}(x)e^{\frac{1}{2}f(\Box)}G_{1\nu}^{(j)d}(x)e^{\frac{1}{2}f(\Box)}G_{2\lambda}^{(j)\nu
eh}(x,y)+&&
\nonumber \\
e^{\frac{1}{2}f(\Box)}G_{3\mu\lambda}^{(j)beh\nu}(x,x,y)e^{\frac{1}{2}f(\Box)}G_{1\nu}^{(j)d}(x)+&& 
\nonumber \\
e^{\frac{1}{2}f(\Box)}G_{2\mu}^{(j)be\nu}(x,x)e^{\frac{1}{2}f(\Box)}G_{2\nu\lambda}^{(j)dh}(x,y)+&&
\nonumber \\
\left.
e^{\frac{1}{2}f(\Box)}G_{3\nu\lambda}^{(j)deh\nu}(x,x,y)e^{\frac{1}{2}f(\Box)}G_{1\mu}^{(j)b}(x)
+e^{\frac{1}{2}f(\Box)}G_{2\nu}^{(j)de\nu}(x,x)e^{\frac{1}{2}f(\Box)}G_{2\mu\lambda}^{(j)bh}(x,y)
\right]-&& \nonumber \\
gf^{abc}e^{\frac{1}{2}f(\Box)}
\left\{
{\bar J}_{2\lambda}^{(\eta,j)bh}(x,y)e^{\frac{1}{2}f(\Box)}\left[\partial_\mu P_1^{(\eta)c}(x)\right]\right.+&& \nonumber \\
\left.{\bar P}_1^{(\eta)b}(x)e^{\frac{1}{2}f(\Box)}\left[\partial_\mu J_{2\lambda}^{(\eta)ch}(x,y)\right]
+\partial_\mu\left[W_{3\lambda}^{(\eta,j)bch}(x,x,y)\right]
\right\}=&& \nonumber \\
e^{\frac{1}{2}f(\Box)}\delta^{ah}\eta_{\mu\lambda}\delta^4(x-y),&&
\end{eqnarray}
where we have introduced the 3P-function
\be
W_{3\lambda}^{(\eta,j)abc}(x,y,z)=Z^{-1}\frac{\delta K_2^{(\eta)ab}(x,y)}{\delta j^{\lambda c}(z)}.
\ee
Turning to the 1P-function for the ghost, We derive it with respect to $\eta^{h}(y)$ to obtain
\bea
&-\Box K_2^{(\eta)ch}(x,y)
-ige^{\frac{1}{2}f^{abc}f(\Box)}L_{2\mu}^{(\eta,j)ah}(x,y)\partial^\mu P_1^{(\eta)b}(x)\nonumber \\
&-igf^{abc}e^{\frac{1}{2}f(\Box)}G_{1\mu}^{(j)a}(x)\partial^\mu K_2^{(\eta)bh}(x,y)
-igf^{abc}\partial^\mu W_{3\mu}^{(\eta,j)abh}(x,x,y) \nonumber \\
&=e^{\frac{1}{2}f(\Box)}\delta^{ch}\delta^4(x-y).
\eea
Here we have introduced the 2P-function
\be
L_{2\mu}^{(\eta,j)ab}(x,y)=\frac{\delta G_1^{(j)a}(x)}{\delta \eta^b(y)}.
\ee
Then, we derive with respect to $j^{h\nu}(y)$. This yields the equation for $J_2$ as
\bea
&-\Box J_2^{(\eta)ch\nu}(x,y) 
-igf^{abc}e^{\frac{1}{2}f(\Box)}G_{2\mu\nu}^{(j)ah}(x,y)\partial^\mu P_1^{(\eta)b}(x) \nonumber \\
&-igf^{abc}e^{\frac{1}{2}f(\Box)}G_{1\mu}^{(j)a}(x)\partial^\mu J_2^{(\eta,j)bh\nu}(x,y) \nonumber \\
&-igf^{abc}\partial^\mu J_{3\mu}^{(\eta,j)abh}(x,x,y)=0,
\eea
where we have introduced the 3P-function
\be
J_{3\mu}^{(\eta,j)abc}(x,y,z)=\frac{\delta J_{2\mu}^{(\eta,j)ab}(x,y)}{\delta j^{c\mu}(z)}.
\ee
Again, setting all the currents to 0, gives
the equations for the 2P-functions as
\bea
\label{eq:G2j0}
\Box G_{2\mu\lambda}^{ah}(x,y)+gf^{abc}
e^{-\frac{1}{2}f(\Box)}\partial^\nu\left[
e^{\frac{1}{2}f(\Box)}G_{3\mu\nu\lambda}^{bch}(x,x,y)+
e^{\frac{1}{2}f(\Box)}G_{2\mu\lambda}^{bh}(x,y)\times
\right.
\nonumber \\
\left.
e^{\frac{1}{2}f(\Box)}G_{1\nu}^{c}(x)+
+e^{\frac{1}{2}f(\Box)}G_{1\mu}^{b}(x)
e^{\frac{1}{2}f(\Box)}G_{2\nu\lambda}^{ch}(x)
\right]- \nonumber \\
gf^{abc}e^{-\frac{1}{2}f(\Box)}
\left[ 
e^{\frac{1}{2}f(\Box)}\partial^\nu G_{2\mu\nu\lambda}^{bch}(x,x,y)+
e^{\frac{1}{2}f(\Box)}\partial^\nu G_{2\mu\lambda}^{bh}(x,y)e^{\frac{1}{2}f(\Box)}G_{1\nu}^{c}(x)+
\right. \nonumber \\
\left.
e^{\frac{1}{2}f(\Box)}\partial^\nu G_{1\mu}^{b}(x)e^{\frac{1}{2}f(\Box)}G_{2\nu\lambda}^{ch}(x,y)
\right]-\nonumber \\
gf^{abc}e^{-\frac{1}{2}f(\Box)}
\left[
e^{\frac{1}{2}f(\Box)}\partial_\mu G_{3\nu\lambda}^{bch\nu}(x,x,y)+
e^{\frac{1}{2}f(\Box)}\partial_\mu G_{2\nu\lambda}^{bh}(x,y)e^{\frac{1}{2}f(\Box)}G_{1}^{c\nu}(x)+
\right.\nonumber \\
\left.
e^{\frac{1}{2}f(\Box)}\partial_\mu G_{1\nu}^{b}(x)e^{\frac{1}{2}f(\Box)}G_{2\lambda}^{ch\nu}(x,y)
\right]+\nonumber \\
g^2f^{abc}f^{cde}e^{-\frac{1}{2}f(\Box)}
\left[
e^{\frac{1}{2}f(\Box)}G_{3\mu\nu\lambda}^{bdh}(x,x,y)e^{\frac{1}{2}f(\Box)}G_1^{\nu e}(x)+
\right.\nonumber \\
e^{\frac{1}{2}f(\Box)}G_{2\mu\nu}^{bd}(x,x)e^{\frac{1}{2}f(\Box)}G_{2\lambda}^{\nu eh}(x,y)
+e^{\frac{1}{2}f(\Box)}\partial^\nu G_{4\mu\nu\lambda}^{bdeh\nu}(x,x,x,y)+
\nonumber \\
e^{\frac{1}{2}f(\Box)}G_{2\mu\lambda}^{bh}(x,y)e^{\frac{1}{2}f(\Box)}G_{1\nu}^{d}(x)e^{\frac{1}{2}f(\Box)}G_1^{\nu
e}(x)+\nonumber \\
e^{\frac{1}{2}f(\Box)}G_{1\mu}^{b}(x)e^{\frac{1}{2}f(\Box)}G_{2\nu\lambda}^{dh}(x,y)e^{\frac{1}{2}f(\Box)}G_1^{\nu
e}(x)+\nonumber \\
e^{\frac{1}{2}f(\Box)}G_{1\mu}^{b}(x)e^{\frac{1}{2}f(\Box)}G_{1\nu}^{d}(x)e^{\frac{1}{2}f(\Box)}G_{2\lambda}^{\nu
eh}(x,y)+\nonumber \\
e^{\frac{1}{2}f(\Box)}G_{3\mu\lambda}^{beh\nu}(x,x,y)e^{\frac{1}{2}f(\Box)}G_{1\nu}^{d}(x)+ 
\nonumber \\
e^{\frac{1}{2}f(\Box)}G_{2\mu}^{be\nu}(x,x)e^{\frac{1}{2}f(\Box)}G_{2\nu\lambda}^{dh}(x,y)+
\nonumber \\
\left.
e^{\frac{1}{2}f(\Box)}G_{3\nu\lambda}^{deh\nu}(x,x,y)e^{\frac{1}{2}f(\Box)}G_{1\mu}^{b}(x)
+e^{\frac{1}{2}f(\Box)}G_{2\nu}^{de\nu}(x,x)e^{\frac{1}{2}f(\Box)}G_{2\mu\lambda}^{bh}(x,y)
\right]-\nonumber \\
gf^{abc}e^{\frac{1}{2}f(\Box)}
\left\{
{\bar J}_{2\lambda}^{bh}(x,y)e^{\frac{1}{2}f(\Box)}\left[\partial_\mu P_1^{c}(x)\right]\right.+
\nonumber \\
\left.{\bar P}_1^{b}(x)e^{\frac{1}{2}f(\Box)}\left[\partial_\mu J_{2\lambda}^{ch}(x,y)\right]
+\partial_\mu\left[W_{3\lambda}^{bch}(x,x,y)\right]
\right\}= \nonumber \\
e^{\frac{1}{2}f(\Box)}\delta^{ah}\eta_{\mu\lambda}\delta^4(x-y).
\eea
For the ghost, one has
\bea
&-\Box K_2^{ch}(x,y)
-ige^{\frac{1}{2}f^{abc}f(\Box)}L_{2\mu}^{ah}(x,y)\partial^\mu P_1^{(\eta)b}(x)\nonumber \\
&-igf^{abc}e^{\frac{1}{2}f(\Box)}G_{1\mu}^{a}(x)\partial^\mu K_2^{bh}(x,y)
-igf^{abc}\partial^\mu W_{3\mu}^{abh}(x,x,y) \nonumber \\
&=e^{\frac{1}{2}f(\Box)}\delta^{ch}\delta^4(x-y),
\eea
and
\bea
&-\Box J_2^{ch\nu}(x,y) 
-igf^{abc}e^{\frac{1}{2}f(\Box)}G_{2\mu\nu}^{ah}(x,y)\partial^\mu P_1^{b}(x) \nonumber \\
&-igf^{abc}e^{\frac{1}{2}f(\Box)}G_{1\mu}^{a}(x)\partial^\mu J_2^{bh\nu}(x,y) \nonumber \\
&-igf^{abc}\partial^\mu J_{3\mu}^{abh}(x,x,y)=0.
\eea

\bibliographystyle{unsrt}

\end{document}